# Electric Field Assisted Charge Transfer Mediated Electronic Coupling in Ferromagnetic/Multiferroic Hetero-structure.

Subho Saha, Sourav Chowdhury, Suryakanta Mondal, Kartick Biswas, Deepak Prajapat, Andrei Gloskovskii, Suresh Bharadwaj, Mukul Gupta, Pavan Nukala, Bhagwati Prasad, Ram Janay Choudhary*

Subho Saha

UGC DAE Consortium for Scientific Research, Indore, India.

Sourav Chowdhury

Email Id: mailatsourav07@gmail.com

Deutsches Elektronen-Synchrotron DESY, Notkestrasse 85, 22607, Hamburg, Germany

Suryakanta Mondal

Department of Materials Engineering, Indian Institute of Science, Bangalore, KA 560012, India.

Email: suryakantam@iisc.ac.in

Kartick Biswas

Center for Nanoscience and Engineering, Indian Institute of Science, Bengaluru, 560012, India

Email: kartickb@iisc.ac.in

Deepak Prajapat

Deutsches Elektronen-Synchrotron DESY, Notkestrasse 85, 22607, Hamburg, Germany

Email Id: deepak.prajapat@desy.de

Andrei Gloskovskii

Deutsches Elektronen-Synchrotron DESY, Notkestrasse 85, 22607, Hamburg, Germany

Email Id: andrei.hloskovsky@desy.de

Suresh Bharadwaj

UGC DAE Consortium for Scientific Research, Indore 452001, India.

Email Id: sb10suresh@gmail.com

Mukul Gupta

UGC DAE Consortium for Scientific Research, Indore 452001, India.

Email: dr.mukul.gupta@gmail.com

Pavan Nukala

Center for Nanoscience and Engineering, Indian Institute of Science, Bengaluru, 560012, India

Email: pnukala@iisc.ac.in

Bhagwati Prasad

Department of Materials Engineering, Indian Institute of Science, Bangalore, KA 560012, India.

Email: bpjoshi@iisc.ac.in

*, R. J. Choudhary*[*]

UGC DAE Consortium for Scientific Research, Indore 452001, India.

**Abstract:**

The strongly coupled ferromagnet-multiferroic material could be a better alternative to complementary metal-oxide semiconductor (CMOS) based data storage device or rather lone multiferroic oxide materials because it offers the electric field controlled conductivity and large magnetisation at the interfaces, but choosing a proper system is still very much challenging. In the present scenario, an interionic charge transfer from Co ion to Fe ion via oxygen is observed at the interfacial region of $LaCoO_3$ (LCO)/$BiFeO_3$ (BFO) grown on $SrTiO_3$ substrate, leading to induce multivalency in Co, Fe ion and also induces exchange biasing in the system. The charge transferring between Co and Fe ions gets further enhanced by electric biasing across the bilayer in the low temperature region, which affects the polarisation behaviour of BFO layer. This confirms the electronic coupling between the ferromagnetic LCO and the multiferroic BFO layers, useful for the making of voltage-controlled magnetic anisotropy (VCMA) systems.

*Corresponding author: ram@csr.res.in

## Introduction:

The manipulation of magnetic and electrical properties at the interface of magneto-electrically coupled hetero-structure has attracted a huge attention due to its richness in fundamental science as well as its potential in the development of cost effective spintronic device [1-3]. In that sense the hetero-structure of multiferroic and ferromagnetic material would be a suitable compound owing to the prominent magnetoelectric coupling between the constituent materials as well as interfacial phenomena such as charge transfer, orbital ordering, octahedral distortion, atomic reconstruction etc. Such magneto-electrically coupled system offers electric field as an additional order parameter in order to engineer the magnetic properties of the interfacial ferromagnetic material. Upon application of the electric field, the space symmetry in the multiferroic materials is broken, which impacts the magnetic exchange interactions at the interface of heterostructure. S.M. Wu et al. revealed the reversal of exchange bias at the LSMO/BFO interface by inducing ferroelectric polarization in the system [4]. Besides, the firm coupling between antiferromagnetic domain and ferroelectric domain in the BFO layer, affects the reversal of exchange bias on the application external electric biasing. However, such hetero-structure/device systems, where electric biasing simultaneously controls the electrical and magnetic properties of the same system remain relatively unexplored. Investigating such systems is of considerable importance, because they hold significant potential for making non-volatile magnetoelectric memory, spintronic devices, magnetically recorded ferroelectric memory devices etc.

BFO is a well-known multiferroic material where antiferromagnetic domains can be determined by imaging the ferroelectric domains in the presence of electrical switching [5-6]. The interfacial exchange coupling in the hetero-structure BFO/LSMO induces a large magnetization in $BiFeO_3$ layer [7]. In BFO/YBCO hetero-structure also, the interfacial exchange coupling gives rise to weak ferromagnetism at the interface [8]. Within these magneto-electrically coupled systems, free electrons are generated at the interface region by applying electric field across the interface. These free charges are spin dependent, which induce additional magnetization in a few nanometres of FM layer from the interface [9]. In such cases, the charge transfer is mediated by electric field. However, there are instances where this charge transfer can also be triggered via substrate induced strain in the bilayer sample. Such strain induced charge transfer is familiar to the interfaces containing LCO, which in its single layer thin film form shows ferromagnetism due to its mixed spin state property [10-11,13]. In most of the bilayer systems, this strain induced charge fluctuation at the interface follows uniaxial direction i.e. either from the top layer to bottom or vice versa. However, modifying the direction of charge fluctuation by applying electric field to the heterostructure and probing the corresponding modifications in the magnetic and electrical property of the bilayer sample is still lacking, which is of huge importance in the field of spintronic device applications. To address this issue, we have taken ferromagnetic LCO as the top layer due to its charge transfer character and multiferroic BFO as a bottom-layer material for its strong magnetoelectric coupling property.

Here we have grown a bilayer heterostructure of LCO/BFO on $SrTiO_3$ (001) (STO) and Niobium doped $SrTiO_3$ (001) (NbSTO) substrates. The heterostructures grown on Nb doped STO are used for carrying out the electric bias dependent studies. The XAS and electrical field dependent photoemission spectroscopy (PES) study unfold the charge transfer from LCO layer to BFO layer. The magnetization measurement reveals an exchange bias effect at the interface.

The electrical measurements reflect that there is a higher leakage current in the bilayer of LCO/BFO compared to the single layer thin film of BFO due to which there is a modification in the polarization of the BFO layer.

**Results and Discussions:**

Figure 1(a) shows the 2θ-ω X-ray diffraction pattern of the bilayer thin film of LCO/BFO grown on STO (001) substrate. The diffraction pattern indicates c-axis growth and the single-phase nature of both LCO and BFO layers on the STO substrate. The calculated lattice parameter in the out-of-plane direction *c* of BFO is 4.048 Å, which is higher than the corresponding bulk value of 3.93 Å, reflecting the tensile strain in the out-of-plane direction. This arises due to the higher pseudocubic lattice parameter of the bulk BFO (3.93 Å), compared to the lattice parameter of the STO substrate (3.90 Å), which would inflict an in-plane compressive strain and out-of-plane tensile strain on the BFO film grown on the STO substrate, as observed. On the other hand, the bulk lattice parameter of LCO is 3.80 Å, which is smaller than the BFO layer. Thus, when the LCO layer is grown on the BFO layer, it will be subjected to in-plane tensile strain and out-of-plane compressive strain. From X-ray diffraction, we obtained an out-of-plane lattice parameter of 3.778 Å, consistent with the out-of-plane compressive strain. Figure 1(b) shows the in-plane reciprocal space mapping (RSM) of the bilayer stack around the STO (103) peak. A prominent spot of the LCO (103) and BFO (103) is observed at the same $Q_x$ value of the STO (103) spot, indicating the highly strained state of the LCO and BFO layers on the STO substrate. The calculated in-plane lattice parameter of LCO and BFO are 3.9Å and 3.89Å respectively. So, LCO grows with 2.6% in-plane tensile strain on STO. Whereas BFO grows under a compressive strain of 1% on STO.  So, X-ray diffraction studies confirm the development of strained epitaxial growth of both the layers on STO.

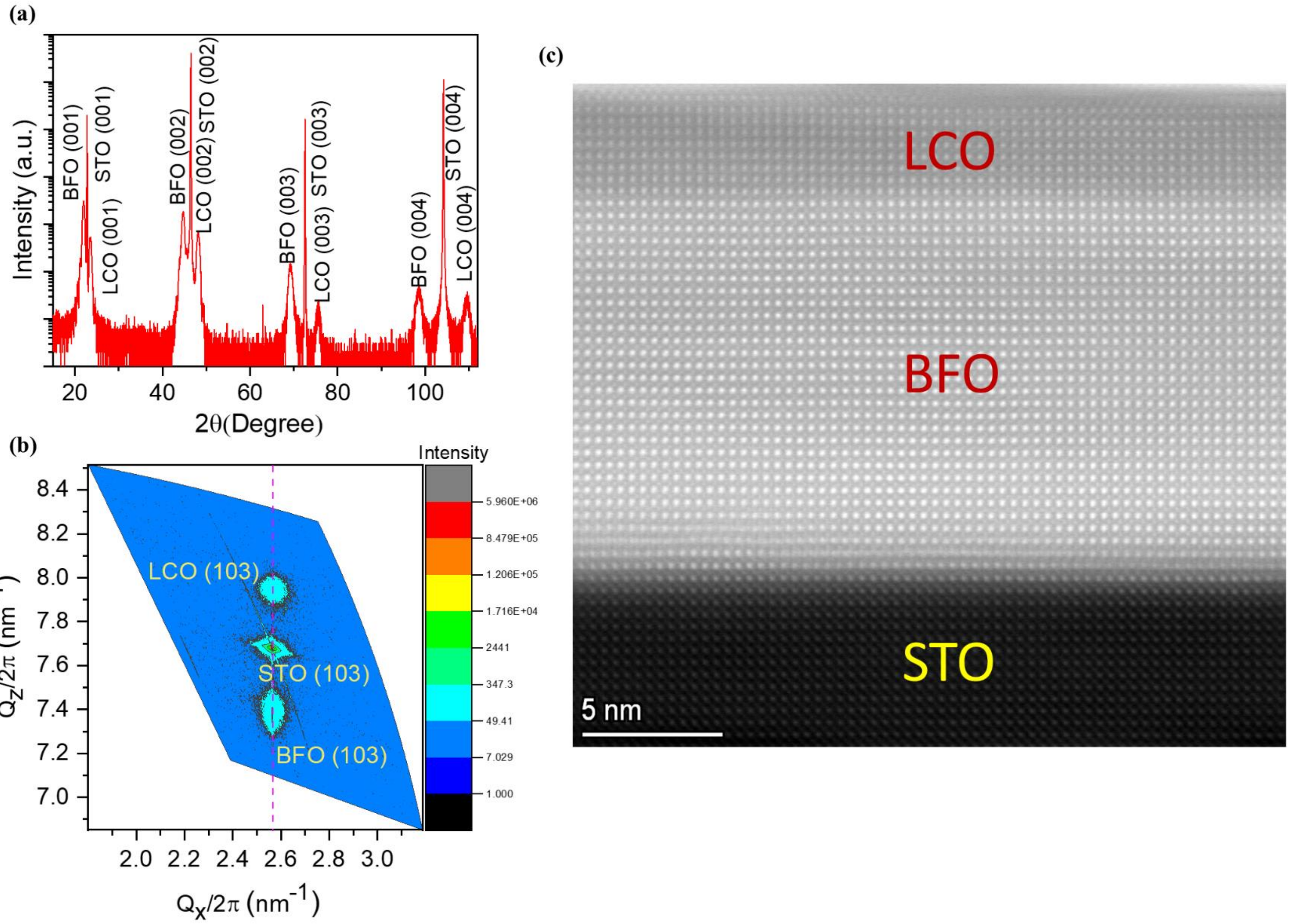

*Fig 1: (a) $2\theta$-$\omega$ X-ray diffraction pattern of the grown bilayer film on STO (001) substrate. (b) Reciprocal space mapping (RSM) plot of the grown bilayer film around the STO (103) plane, revealing the highly strained in-plane growth. (c) High Resolution Transmission Electron Microscopy (HRTEM) images showing the individual layers LCO and BFO grown on STO(001) substrate.*

Cross sectional Transmission Electron Microscopy (TEM) images reflect that both the layers LCO (~ 5 nm) and BFO (~ 15 nm) of the hetero-structure grow coherently on the STO substrate. The sharp interface between the LCO and BFO layers, indicates that there is no any inter diffusion occurring at the interface of LCO and BFO heterostructure as shown in figure 1(c).

To find out the electronic states of Co and Fe ions in the bilayer, we have performed the soft XAS measurement at Co and Fe L-edge of the bilayer [figure 2(a) and (b)]. In the Co or Fe L-edge spectra, the excitation of the electrons from the occupied state of Co or Fe 2p level to its unoccupied state of 3d level ($2p^6d^n \rightarrow 2p^5d^{n+1}$) creates the entire Co or Fe L-edge spectrum [11-14]. In the Co L-edge spectra of LCO film, the appearance of $L_3$ and $L_2$ features positioned at 781 eV and 796 eV reveals the presence of high spin state of Co 3+ charge state. The shoulder feature at the right side of $L_3$ edge indicates the existence of low spin state of Co 3+ charge state as well in the LCO film [13]. So in LCO film, $Co^{3+}$ is present with both low and high spin states. These observations are consistent with the previous reports [10,13]. In the Co L-edge spectrum of the bilayer LCO/BFO, it is observed that the peak position of both $L_3$ and $L_2$ features is shifted by 0.6 eV towards the higher energy side in the spectrum, as compared to the single layer of LCO [figure 2(a)]. This indicates the appearance of higher charge state of Co ions in the LCO/BFO bilayer compared to the single layer of LCO, wherein Co stays only in 3+ charge state. So at the interface of the heterostructure, some amount of Co is also present in the 4+ charge state along with the 3+ charge state [15]. The creation of this multi-valency in Co at the interface of LCO and BFO bilayer indicates the possibility of interionic charge transfer between the transition metals via oxygen.

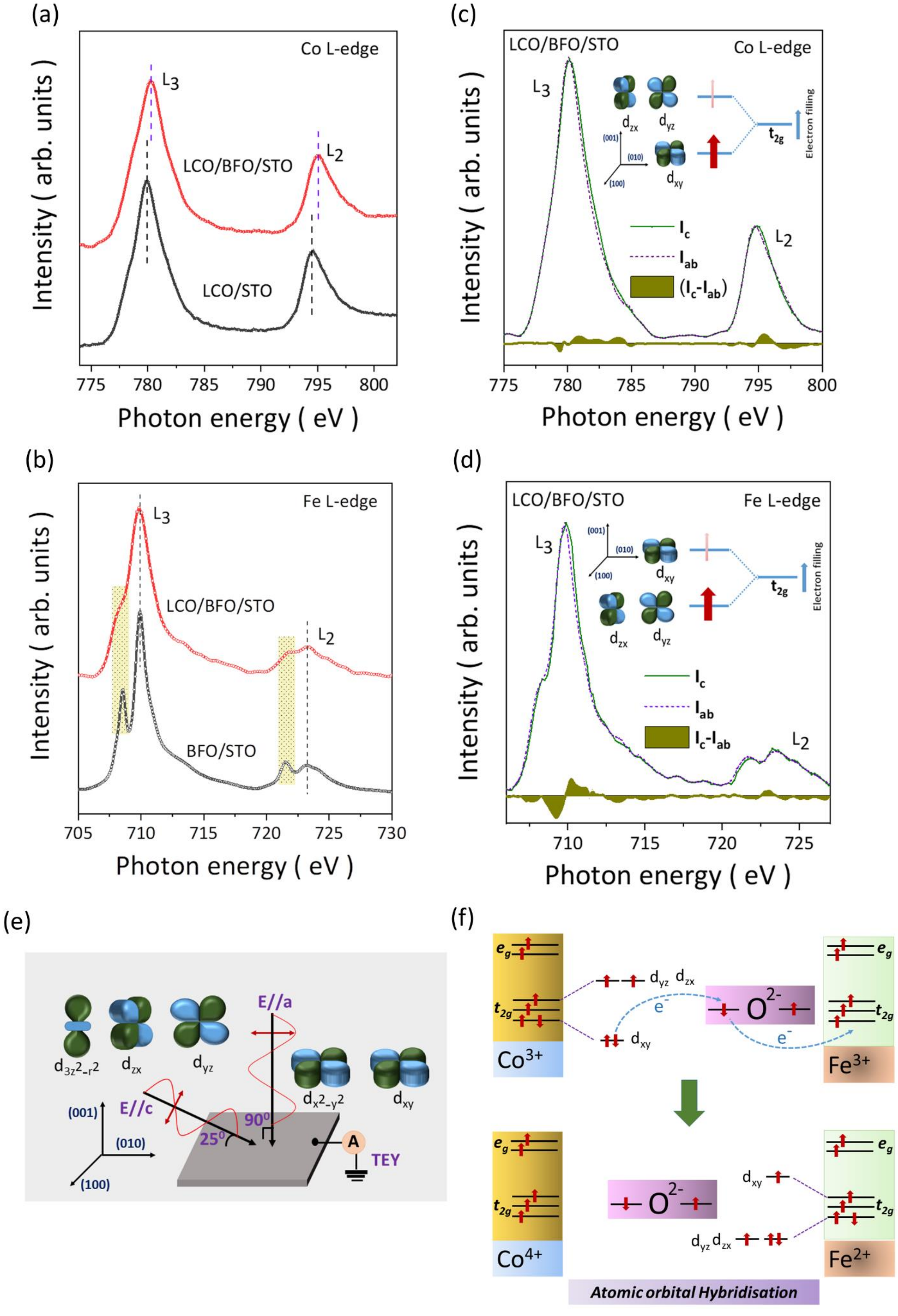

(a)
Co L-edge
L3
L2
LCO/BFO/STO
LCO/STO
Intensity ( arb. units )
Photon energy ( eV )
775 780 785 790 795 800
(b)
Fe L-edge
L3
L2
LCO/BFO/STO
BFO/STO
Intensity ( arb. units )
Photon energy ( eV )
705 710 715 720 725 730
(c)
LCO/BFO/STO
Co L-edge
L3
L2
dzx dyz dxy
(001) (010) (100)
t2g
Electron filling
Ic
Iab
(Ic-Iab)
Intensity ( arb. units )
Photon energy ( eV )
775 780 785 790 795 800
(d)
LCO/BFO/STO
Fe L-edge
L3
L2
(001) (010) (100)
dxy dzx dyz
t2g
Electron filling
Ic
Iab
Ic-Iab
Intensity ( arb. units )
Photon energy ( eV )
710 715 720 725
(e)
d3z2-r2 dzx dyz dx2-y2 dxy
E//a
E//c
90°
25°
(001) (010) (100)
A
TEY
(f)
eg
t2g
Co3+
dyz dzx
dxy
O2-
e-
e-
eg
t2g
Fe3+
eg
t2g
Co4+
O2-
dxy
dyz dzx
eg
t2g
Fe2+
Atomic orbital Hybridisation

*Figure 2. Electronic study: (a) and (b) experimental spectra of Co L-edge and Fe L-edge of heterostructure LCO/BFO/STO and single layer LCO/STO, BFO/STO (as a reference). Orbital polarization analysis: (c) and (d) Polarization dependent XLD signal of LCO/BFO/STO heterostructure at Co L-edge and Fe L-edge. The Co $d_{xy}$ and Fe $d_{yz}$ / $d_{yz}$ orbital polarization is shown schematically in the inset, respectively. (e) Experimental geometry of the Polarization dependent XAS measurements shown schematically. (f) a schematic diagram that briefs the interionic charge transfer from Co ion to Fe ion via oxygen and the corresponding orbital hybridisation.*

To validate the interionic charge transfer between the transition metals (Co and Fe) at the interface of LCO and BFO layer, we have carried out XAS measurement at Fe $L_{3,2}$ edge also of the bilayer film along with the single layer thin film of BFO as shown in figure 2(b) [16-18]. The spin orbit coupling splits Fe L-edge into two peaks $L_3$ and $L_2$, which show some fine features due to octahedral crystal field splited $t_{2g}$ and $e_g$ states [19-20]. Interestingly, the separation between $t_{2g}$ and $e_g$ is not so visible in the Fe L-edge for the bilayer sample, which indicates a modified electron occupancy in the orbitals ($e_g$ and $t_{2g}$) as compared to the film of BFO. A closer look at the left shoulder side of both the $L_3$ and $L_2$ edges reveals the decrease in the intensity of $t_{2g}$ states, as compared to the single layer of BFO [21]. This reduction in the intensity of $t_{2g}$ level indicates the lesser number of unoccupied state, which can occur when there is filling of those unoccupied orbitals in the bilayer. The XAS of Co L-edge of the bilayer revealed the presence of $Co^{4+}$ states also besides the $Co^{3+}$ state. The appearance of $Co^{4+}$ state and decreased intensity of $t_{2g}$ feature in the Fe L-edge suggest that $Co^{3+}$ in then LCO layer donates an electron to the ligand (Oxygen) and the extra electron from the ligand site is transferred to Fe ions in the BFO layer, occupying its $t_{2g}$ state. As a result of which some amount of Fe ion's charge state gets converted from 3+ state to 2+ state. So the XAS study of Co and Fe L-edge indicates the occurrence of an interionic charge transfer via ligand (Oxygen) at the interface between LCO and BFO layer.

To further probe the electronic occupation in the orbitals of the transition metals, we have carried out X-ray linear dichroism measurement of the heterostructure as shown in figure 2(c-d). To get the dichroic signal, we have shined the surface of the thin film by photons in the normal incidence ($I_{ab}$) and grazing incidence ($I_c$) modes. It should be noted that the polarization direction of the incident photon governs the absorption of electrons in a particular orbit, in a way that if the XLD signal (i.e. $I_c - I_{ab} > 0$) is positive, it indicates the in-plane orbital polarization. However, if the signal ($I_c - I_{ab} < 0$) is negative, it signifies the out of plane orbital polarization [22-23]. Figure 2(c) revealed the positive XLD signal for Co L-edge of the bilayer, suggesting that the Co 3d electrons prefer to occupy in the in-plane orbitals [24]. The XAS study reveals Co has two charge states 3+ and 4+ having electronic configuration $t_{2g}^4 e_g^2$ and $t_{2g}^3 e_g^2$ in high spin state. Out of these two configurations, effectively the $t_{2g}^4 e_g^2$ can show dichroic signal due its partial filling nature, where initially $t_{2g}$ and $e_g$ orbitals get half-filled with the electrons ($t_{2g}^3 e_g^2$) and then the last electron breaks the degeneracy between orbitals $t_{xy}$ and $t_{xz}$, $t_{yz}$ by filling up into the $t_{xy}$ orbital and hence we obtained in-plane orbital polarization from the Co L-edge XLD analysis. The XLD signal from Fe L-edge is negative i.e. $I_c - I_{ab} < 0$, which indicates out of plane polarization shown in figure 2(d) [25]. XAS study establishes two charge states 3+ and 2+ for Fe which has electronic configuration $t_{2g}^3e_g^2$ and $t_{2g}^4e_g^2$. Out of these, as for the Co L-edge, the partially filled $t_{2g}^4e_g^2$ will yield significant dichroic signal.

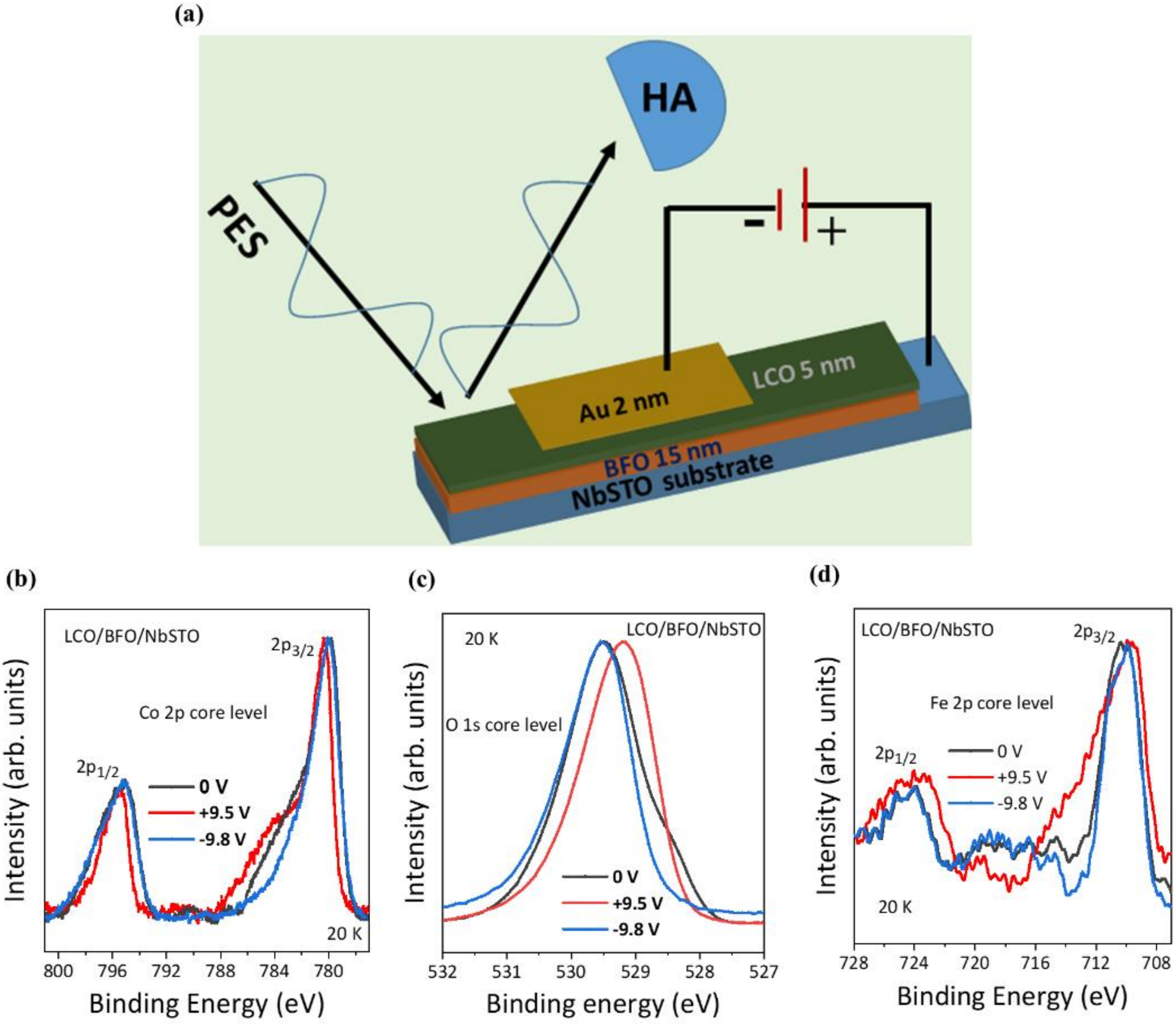

*Figure 3. Electrical analysis: (a) Schematic experimental geometry for the electric field dependent PES measurements. (b) Co 2p, (c) Oxygen 1s and (d) Fe 2p core level spectra obtained under the application of different biasing voltage ( 0 V, +9.5 V and -9.8 V ) at temperature 20 K.*

To further understand the direction of this charge transfer, we have carried out electric field dependent photoemission spectroscopy measurement on the bilayer of LCO/BFO grown on Nb doped STO (001) substrate at temperature 20 K [figure 3(a-d)]. We have recorded the Co 2p,

O 1s and Fe 2p core level spectra in order to find out the change in their charge state while applying different mode of biasing across the bilayer sample as shown in figure 3(b-d). Two prominent peaks were observed in Co 2p core level, which represent the spin orbit split state of Co $2p_{3/2}$ and Co $2p_{1/2}$ in the zero biasing state. The binding energy positions of the Co $2p_{3/2}$ and $2p_{1/2}$ state are observed to be 779.9 eV and 794.8 eV respectively and these are followed by their satellite states [figure 3(b)]. The best fitting of this PE spectra reflects the presence of Co 3+ and 4+ charge states in the system, as shown in figure S2(a). This study of PES at zero biasing, which shows the multiple valency Co (3+ and 4+) is consistent with the XAS analysis of Co L-edge. Interestingly, when a positive biasing of 9.5 volt is applied across the bilayer, the corresponding binding energy position is slightly shifted to higher values of 780.3 eV and 795.2 eV respectively [figure 3(b)] [26-27,38]. The best fitting of this PE spectra shows that there is an increment and decrement in the percentage of Co 4+ and 3+ charge states respectively [figure S2(b)]. The change in the partial amount of two different charge states and a slight shift in $L_3$ and $L_2$ edge towards higher binding energy values indicate that Co which initially possessed dominant of 3+ charge state compared to 4+ charge state, upon electric biasing has dominant 4+ charge state compared to 3+ charge state. This shows that there is an additional charge flow from Co ion towards the ligand when positive biasing is applied to the system. Surprisingly, there is no significant change when negative biasing is applied to the same sample, which indicates that the transfer of charge (electron) follows uniaxial direction i.e. from Co ion to Oxygen.

In order to validate the uniaxial transport of electron, we have also recorded Oxygen 1s core level spectra by applying different mode of biasing across the bilayer sample [figure 3(c)]. The binding energy position of O1s is 529.5 eV in the presence of zero biasing, which is shifted towards lower binding energy position at 529.2 eV, when positive biasing is applied to the system. This reflects the appearance of the extra electrons at oxygen sites due to which we need lesser energy to remove an electron from oxygen 1s core level. But no change is observed in the binding energy of oxygen 1s when negative biasing is applied to the bilayer sample. Again, two prominent peaks are observed in the Fe 2p core level spectra, corresponding to Fe $2p_{3/2}$ and Fe $2p_{1/2}$ states positioned at 710.1 eV and 724.3 eV under the application of zero biasing, respectively, as shown in figure 3(d). But, these positions shifted towards the lower binding energy side when a positive biasing is applied across the system. This indicates the further induction Fe 2+ charge state from Fe 3+ charge state under the application of positive biasing into the system [27-28]. Certainly, the B.E. position of Fe $2p_{3/2}$ and Fe $2p_{1/2}$ state remain unchanged under the application of negative biasing. Thus, this electric field dependent PES study reveals that the electron transferred from Co ion in LCO layer appears at the Fe site in BFO layer via oxygen, when positive electric biasing is applied across the bilayer sample at 20 K.

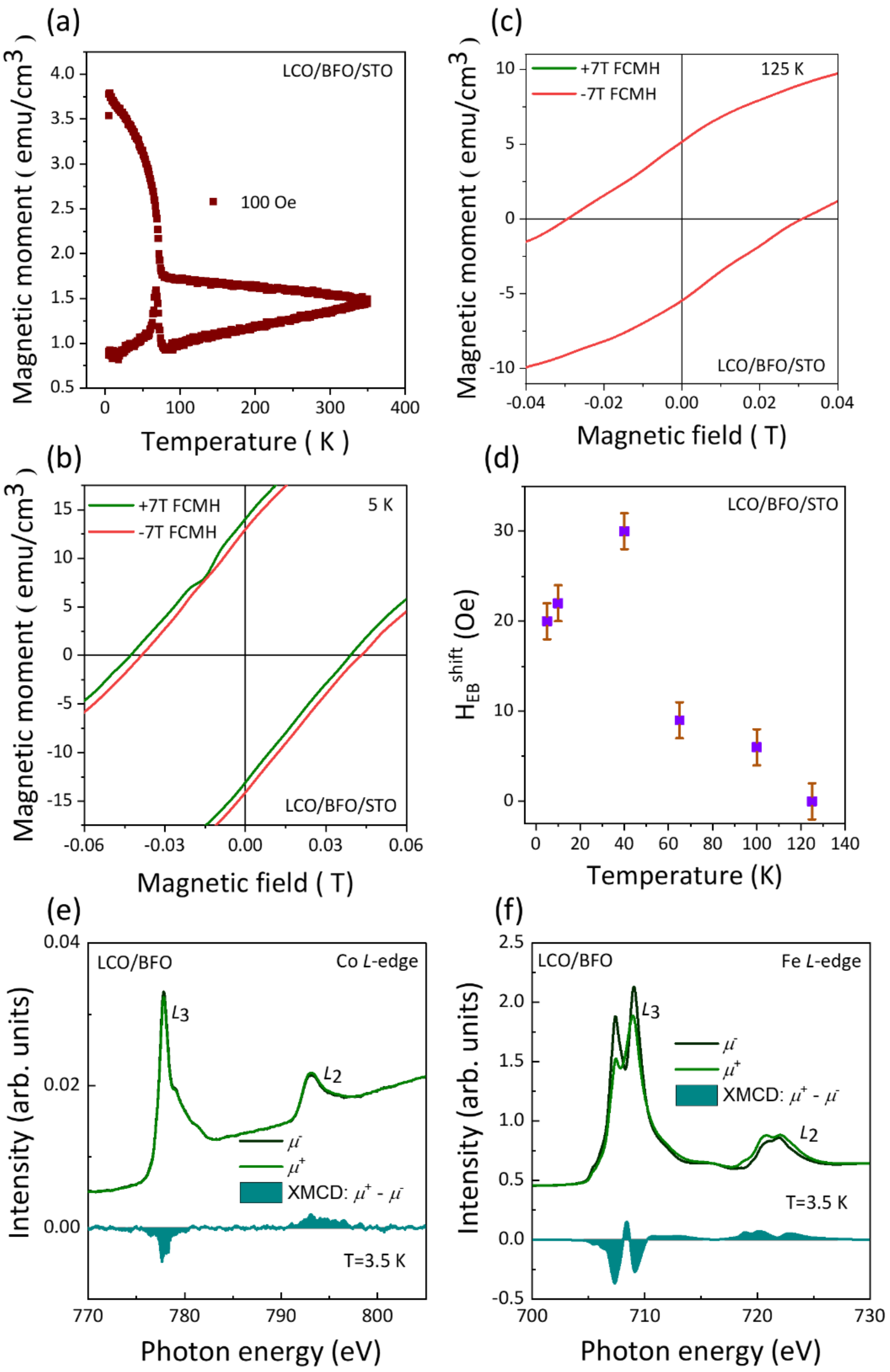
(a)
LCO/BFO/STO
100 Oe
Magnetic moment ( emu/cm$^3$ )
Temperature ( K )
(c)
+7T FCMH
-7T FCMH
125 K
LCO/BFO/STO
Magnetic moment ( emu/cm$^3$ )
Magnetic field ( T)
(b)
+7T FCMH
-7T FCMH
5 K
LCO/BFO/STO
Magnetic moment ( emu/cm$^3$ )
Magnetic field ( T)
(d)
LCO/BFO/STO
$H_{EB}^{shift}$ (Oe)
Temperature (K)
(e)
LCO/BFO
Co L-edge
$L_3$
$L_2$
$\mu^-$
$\mu^+$
XMCD: $\mu^+$ - $\mu^-$
T=3.5 K
Intensity (arb. units)
Photon energy (eV)
(f)
LCO/BFO
Fe L-edge
$L_3$
$L_2$
$\mu^-$
$\mu^+$
XMCD: $\mu^+$ - $\mu^-$
T=3.5 K
Intensity (arb. units)
Photon energy (eV)

*Figure 4. Magnetisation study: (a) Magnetic moment v/s temperature curve at zero field cooled and field cooled warming cycles under the application of 100 Oe magnetic field. (b) and (c) M-H plot taken at FC cycles at a temperature of 5 K and 125 K. (d) $H_{EB}^{shift}$ v/s Temperature curve for the film LCO/BFO/STO. XMCD analysis of (e) Co L-edge and (f) Fe L-edge for the film LCO/BFO/STO at a temperature of 3.5 K.*

The figure 4(a) shows magnetization behaviour of the bilayer sample, which reveals paramagnetic to ferromagnetic transition below 76 K [13]. This transition is due to the LCO layer of the bilayer, since similar transition is also observed in the LCO film, as shown in figure [S4(a)]. In the single layer of BFO, a spin reorientation is observed around 220 K as shown in figure [S4(b)] [28-29]. But this spin reorientation is absent in the bilayer sample, owing to a strong coupling at the interface between the individual layer of LCO and BFO. To further understand the magnetic interaction between the layers of the bilayer, we have recorded the positive and negative field cooled M-H measurement of the bilayer sample in the magnetic field range of ±7 T at temperature 5 K as shown in figure 4(b). A clear hysteresis loop is obtained at temperature 5 K, which is due to the ferromagnetic nature of LCO layer below 76 K [figure S5(a)]. A shift in the hysteresis loop is also observed along the horizontal axis, which reveals the magnetic exchange bias effect (MEBE) at the interface of the bilayer sample. The characteristic value of the exchange bias field shift can be calculated from the given formula: $H_{EB}^{shift} = (H_+ + H_-)/2$, where $H_+$ and $H_-$ are the coercive field values in the positive and negative cycles of the hysteresis loop. The value of coercive field for the hysteresis loop taken at 5 K for the film f1 are $H_+ = +430$ Oe and $H_- = -390$ Oe and $H_{EB}^{shift}$ is estimated to be around 20 Oe**.** To find out the blocking temperature, we have carried out the M-H loop by recording field cooled +7 T and -7 T cycles at temperatures 10 K, 40 K, 65 K, 100 K, 125 K and acquired the value of blocking temperature to be 125 K as shown in figure S5(a-f) for the film.

To probe the magnetic moment alignment of the transition metal elements at the interfacial region between LCO and BFO, we have carried out X-ray magnetic circular dichroism (XMCD) measurements at the Co and Fe L-edge sites of the heterostructure film LCO/BFO/STO at 3.5 K. The XMCD was obtained by taking the difference of the XAS spectra of right circularly polarized light ($\mu^+$) and left circularly polarized light ($\mu^-$), i.e., $\mu^+ - \mu^-$, by flipping the X-ray helicity, at a fixed magnetic field of 6 T. The XMCD signal at the Co L-edge reveals a negative value at the $L_3$ edge and a positive signal at the $L_2$ edge, suggesting ferromagnetic alignment of Co moments (figure 4(e)) [30]. Like Co ions, the XMCD signal at the Fe L-edge is also negative at the $L_3$ edge and a positive signal at the $L_2$ edge, indicating the ferromagnetic alignment Fe moments as well (figure 4(f))[31]. As discussed above, the XAS measurements reflected the presence of hetero valency of Co (3+ and 4+) and Fe (3+ and 2+) ions in the bilayer thin film of LCO/BFO. These hetero-valence states of Co and Fe ion will induce a ferromagnetic double exchange interaction via oxygen, as revealed by the XMCD results. The sign of the XMCD signal is similar for the Co and Fe sites, which divulges the parallel alignment of the magnetic moments of Co and Fe ions (figure 4(e) and (f)).

As discussed above, XAS analysis revealed that Co and Fe ions in the bilayer exhibit multivalent charge states (3+, 4+ and 3+, 2+) respectively, resulting from interionic charge transfer at the heterostructure interface. XLD analysis showed that both Co and Fe ions, with different charge states, reside in high-spin states. Accordingly, the electronic configuration of high-spin $Co^{4+}$ state is $t_{2g}^3e_g^2$, and that of high-spin $Fe^{2+}$ state is $t_{2g}^4e_g^2$. According to the GKA rule, the effective magnetic exchange interaction between $Co^{4+}$ and $Fe^{2+}$ via oxygen is ferromagnetic. Similarly, the superexchange interactions of $Co^{3+}$-O-$Co^{3+}$, $Co^{4+}$-O-$Co^{4+}$ and $Fe^{3+}$-O-$Fe^{3+}$, $Fe^{2+}$-O-$Fe^{2+}$ will give rise to AFM ordering into the system. The exchange interactions $Co^{3+}$-O-$Co^{4+}$ , $Fe^{3+}$-O-$Fe^{2+}$ , $Co^{4+}$-O-$Fe^{2+}$ cause FM ordering and $Co^{3+}$-O-$Co^{3+}$, $Co^{4+}$-O-$Co^{4+}$ and $Fe^{3+}$-O-$Fe^{3+}$, $Fe^{2+}$-O-$Fe^{2+}$ cause AFM ordering [32-34]. These competing interactions at and across the interface pin the magnetic moment causing the magnetic exchange bias effect (MEBE) in the system.

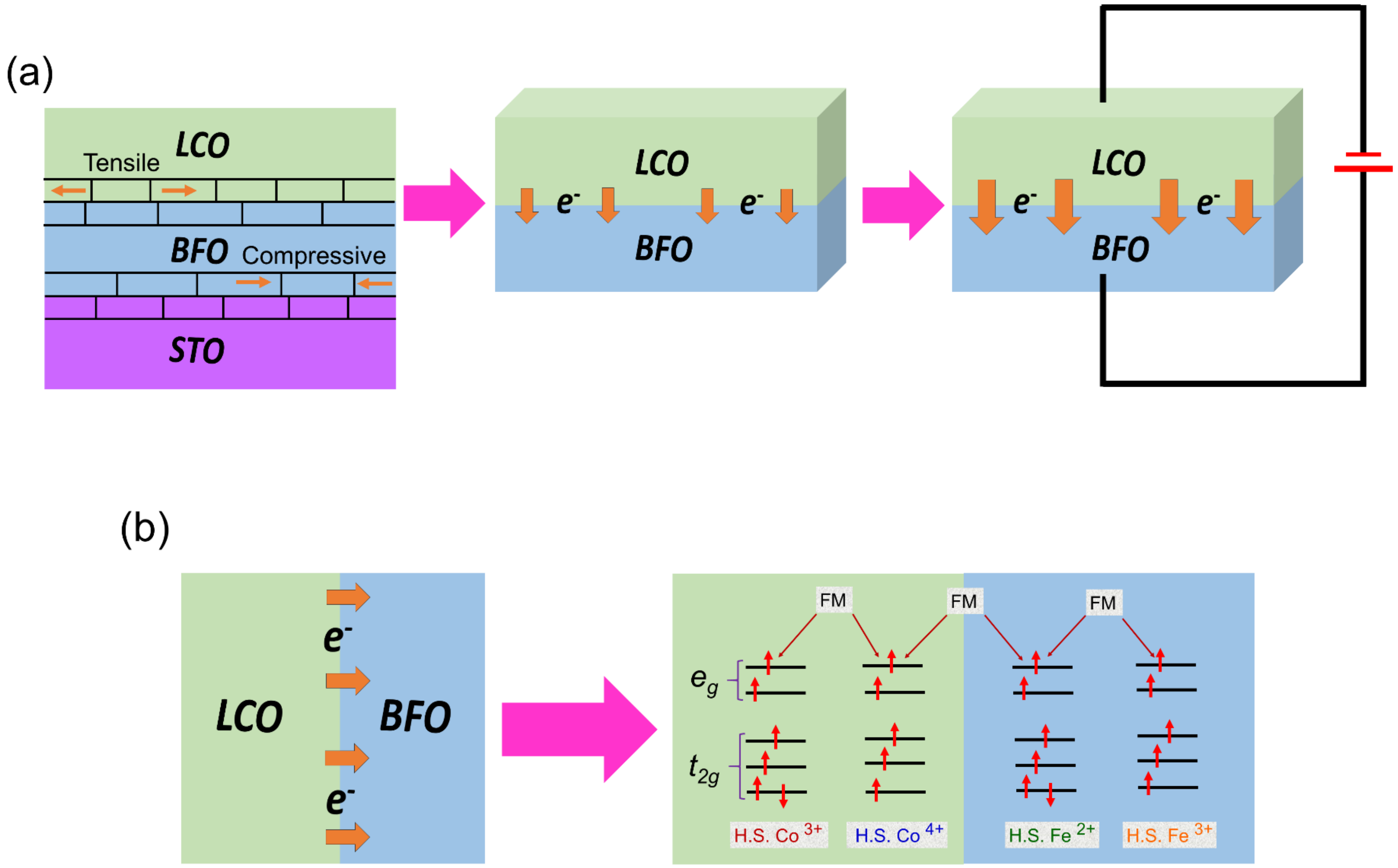

*Figure 5. Microscopic view of (a) the electronic charge transfers due to strain and electric biasing and (b) the magnetic exchange interactions occurring at the interface of the bilayer of LCO/BFO grown on STO (001) substrate.*

As discussed above, we have a strained bilayer of LCO/BFO on STO substrate, which invokes a strong hybridisation between the transition metals (Co and Fe) via ligand (oxygen) and initiates a charge transfer from Co ion to Fe ion via oxygen. This causes an electronic instability at the BFO site and hence we obtained lesser remnant polarisation in the bilayer LCO/BFO compared to the single layer of BFO at room temperature. Earlier study reveals that the strain state in the single layer thin film of BFO gets enhanced due to its piezoelectric behaviour [35-36]. Its lattice parameter gets altered by applying an external electric field across BFO thin films further affecting the strain state. So here, when we have applied an external electric field across the bilayer sample LCO/BFO/NbSTO, the strain state in the bilayer might have enhanced due to which we obtained more charge transfer from Co site to Fe (as confirmed from PES measurement) at 20 K. It is interesting to note that Co changes its spin state from high value to low value with the decrease in the temperature for the single layer thin film of LCO [ 13,37 ]. However, in the present study, when we apply an electric biasing across the bilayer, it provides an extra potential to the electrons of $Co^{3+}$ site ($3d^6$), which felicitates its hopping to Fe site via oxygen rather than arranging into intermediate or low spin state of Co at low temperature. As a result, $Fe^{2+}$ concentration enhances in BFO layer, leading to enhanced electronic hopping between $Fe^{3+}$ and $Fe^{2+}$ in BFO. This results in the decrement in the value of remnant polarisation of the bilayer sample measured at temperature 30 K, compared to the remnant polarisation of the very same bilayer measured at room temperature.

Conclusion:

In summary, LCO/BFO heterostructure is grown on $SrTiO_3$ substrate to probe the charge transfer phenomena across the interface. It divulges uniaxial direction of charge transfer from LCO layer to BFO layer across the interface of the strained heterostructure at room temperature, which gets further elevated by the electric biasing at low temperature region. The unidirectional charge transfer at the interface produces $Fe^{2+}$ state, which disturbs the electronic stability in the BFO layer of the bilayer sample leading to decrement in the polarisation compared to the single layer BFO thin film. The charge transfer phenomena is further enhanced in the presence of electric field biasing across the interface of LCO/BFO at low temperature leading to enhanced $Fe^{2+}$ proportion as well as further decrement of polarisation of the bilayer system. The charge transfer induced multiple valency at Co (3+ and 4+ charge state) and Fe (3+ and 2+ charge state) ions induces both ferromagnetic as well as antiferromagnetic exchange interactions between the transition metals via oxygen, which leads to the formation of magnetic exchange bias effect at the interface of bilayer sample. These modulations in the electrical and magnetic character at the interface between the LCO and BFO layers, affirm their possible magnetoelectric coupling, which will be useful in the making of data storage and spintronic devices.

**Acknowledgements**

The authors are thankful to Avinash Wadikar, Rakesh Sah for the XAS and XLD measurements. Suryakanta Mondal and Bhagwati Prasad are grateful for the departmental XRD facilities of the Department of Materials Engineering at IISc, Bangalore. The authors acknowledge the Micro and Nano Characterization Facility (MNCF), CeNSE, IISc, for TEM sample preparation and the Advanced Facility for Microscopy and Microanalysis (AFMM), IISc, for HRSTEM imaging. Authors are also thankful to Satish Yadav, Binoy Dey for the assistance in experimental measurements.

**Data Availability Statement**

The data that supports the findings of this study are available within the article.

**Table of Contents**

An interionic charge transfer occurred at the interface of highly strained bilayer $LaCoO_3$(LCO)/ $BiFeO_3$ (BFO) grown on $SrTiO_3$(001) substrate. This charge transfer, gets enhanced when electric biasing is applying across the bilayer and also induces multivalency in transition metals, generates novel exchange bias effect. This study reveals a new electronically coupled Ferromagnet/Multiferroic system, useful for logic and magnetic data storage device applications.

**Tittle- Electric Field Assisted Charge Transfer Mediated Electronic Coupling in Ferromagnetic/Multiferroic Hetero-structure.**

**ToC Figure**

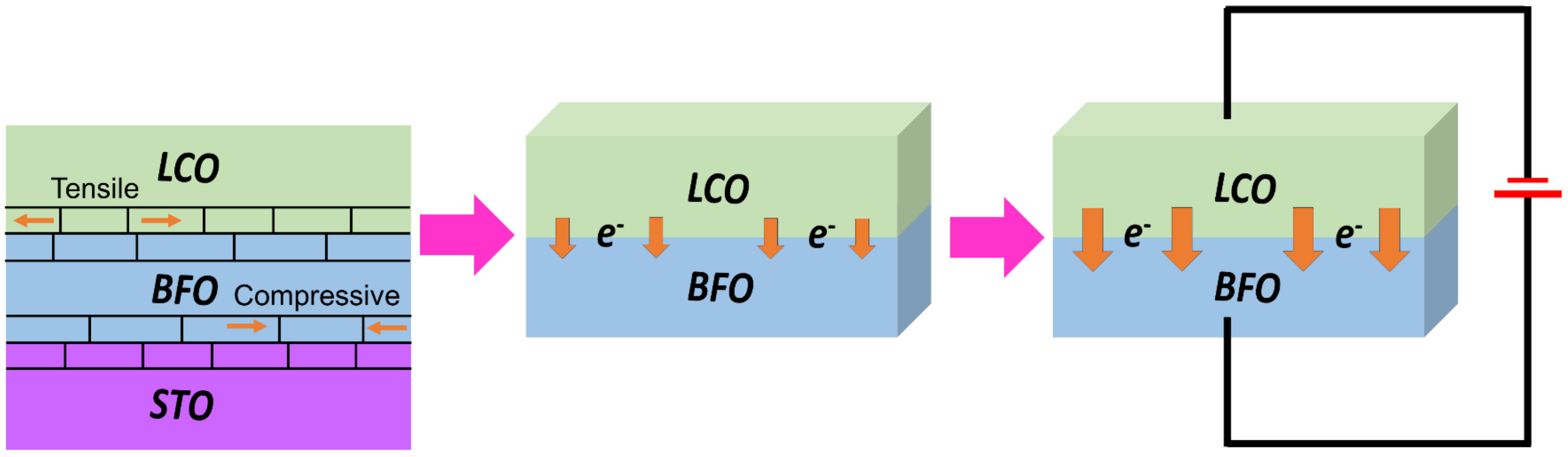

**Supporting information:**

**Tittle: Electric Field Assisted Charge Transfer Mediated Electronic Coupling in Ferromagnetic/Multiferroic Hetero-structure.**

Subho Saha[1], Sourav Chowdhury[2], Suryakanta Mondal[3], Kartick Biswas[4], Deepak Prajapat[2], Andrei Gloskovskii[2],Suresh Bharadwaj[1], Mukul Gupta[1], Pavan Nukala[4], Bhagwati Prasad[3], Ram Janay Choudhary[1*]

UGC DAE Consortium for Scientific Research, Indore, India

Deutsches Elektronen-Synchrotron DESY, Notkestrasse 85, 22607, Hamburg, Germany

Department of Materials Engineering, Indian Institute of Science, Bangalore, KA 560012, India.

Center for Nanoscience and Engineering, Indian Institute of Science, Bengaluru, 560012, India

**Structural Measurements:** The X-ray reflectivity measurement of bilayer LCO/BFO heterostructure on STO (001) substrate figure S1 (a) shows the thickness of the individual layer of LCO and BFO is 5 nm and 15 nm as shown in figure S1(a).

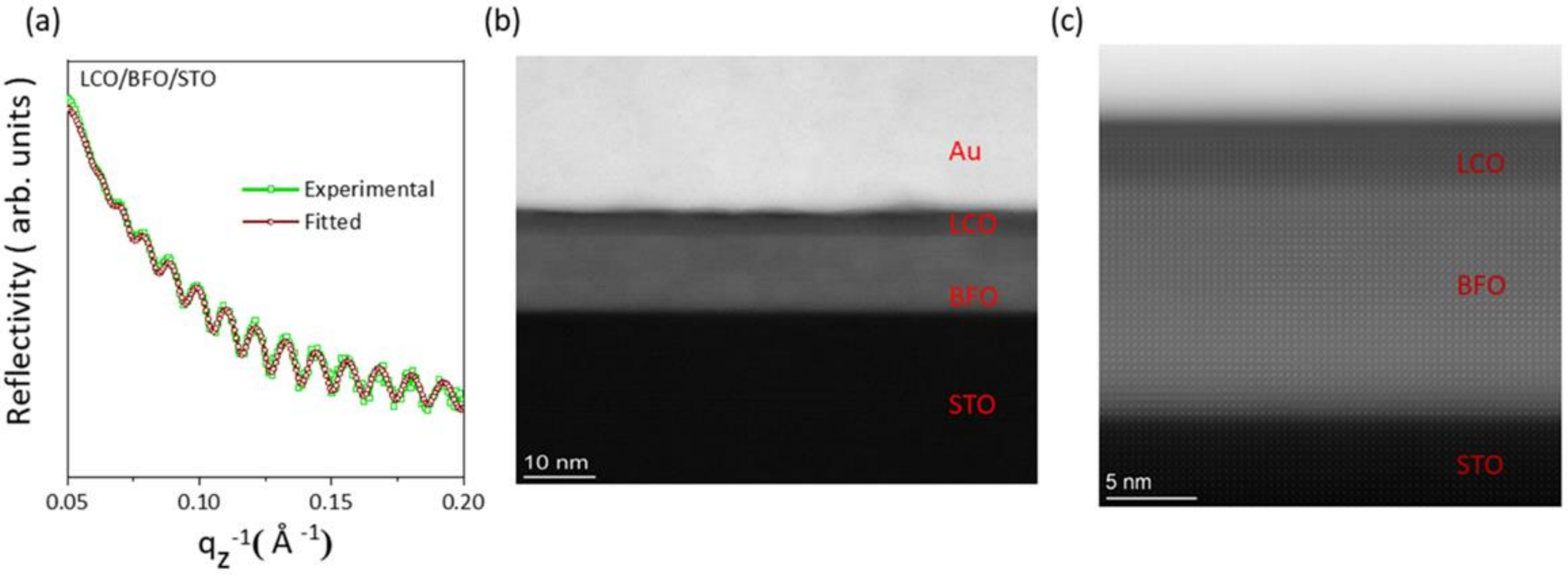

*Figure S1. (a) X-ray reflectivity study of bilayer LCO/BFO grown on STO(001) substrate. (b-c) High Resolution Transmission Electron Microscopy (HRTEM) images showing the individual layers LCO and BFO grown on STO(001) substrate.*

**Electrical Measurements:**

To probe heterovalency of transition metals (Co and Fe) at the interfacial region, we obtained Co 2p and Fe 2p core level spectra in the presence/absence of electric biasing across LCO/BFO/NbSTO at temperature 20 K. Both Co 2p and Fe 2p core level spectra reveal two main peaks corresponding to $2p_{3/2}$ and $2p_{1/2}$ states, which are accompanied by their respective satellite states, as shown in figure S2(a-d) [1-3]. The details of the peak positions of the convoluted features along with their intensity for Co and Fe ion is provided in the Table-1.

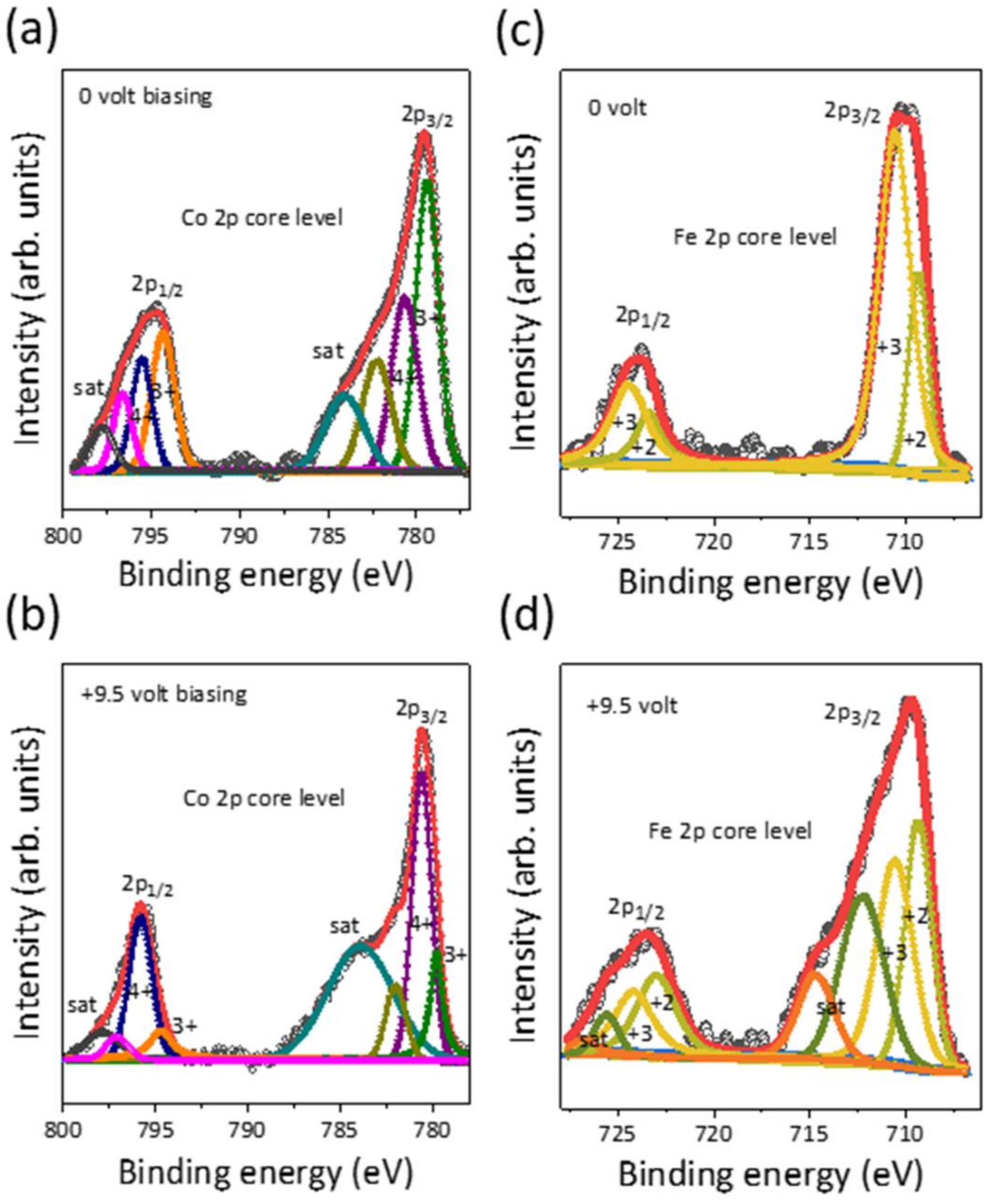

*Figure S2. (a-b) Deconvoluted Co 2p and (c-d) Fe 2p PE spectra of LCO/BFO/NbSTO film at zero and +9.5 volt biasing at temperature 20 K, respectively.*

Table-1

| Elements | Chemical state | Feature no. | Binding energy (eV) position | Relative area(%) V=0 V | Relative area(%) V=+9.5 V |
|---|---|---|---|---|---|
| Co | $Co^{3+}$ $2p_{3/2}$ | 1 | 779.6 | 26 | 10 |
| | $Co^{4+}$ $2p_{3/2}$ | 2 | 780.6 | 18 | 26 |
| | $Co^{3+}$ $2p_{3/2}$ sat | 3 | 782.1 | 12 | 7 |
| | $Co^{4+}$ $2p_{3/2}$ sat | 4 | 784.0 | 11 | 32 |
| | $Co^{3+}$ $2p_{1/2}$ | 5 | 794.5 | 13 | 5 |
| | $Co^{4+}$ $2p_{1/2}$ | 6 | 795.6 | 10 | 14 |
| | $Co^{3+}$ $2p_{1/2}$ sat | 7 | 797.0 | 6 | 2 |
| | $Co^{4+}2p_{1/2}$sat | 8 | 797.8 | 4 | 4 |
| Fe | $Fe^{2+}$ $2p_{3/2}$ | 1 | 709.3 | 17 | 20 |
| | $Fe^{3+}$ $2p_{3/2}$ | 2 | 710.5 | 48 | 23 |
| | $Fe^{2+}$ $2p_{3/2}$ sat | 3 | 712.2 | - | 21 |
| | $Fe^{3+}$ $2p_{3/2}$ sat | 4 | 714.7 | - | 9 |
| | $Fe^{2+}$ $2p_{1/2}$ | 5 | 723.2 | 9 | 10 |
| | $Fe^{3+}$ $2p_{1/2}$ | 6 | 724.3 | 26 | 11 |
| | $Fe^{2+}$ $2p_{1/2}$ sat | 7 | 725.6 | - | 4 |
| | $Fe^{3+}2p_{1/2}$sat | 8 | 726.8 | - | 2 |

We have carried out electrical measurement on the bilayer sample and compared it with the single layer of BFO film. Figure S3(a) illustrates the variation of leakage current with respect to the applied voltage/electric field for BFO single layer and LCO/BFO bilayer grown on Nb doped STO (001) substrate. The leakage current for the single layer of BFO is lesser compared to the bilayer of LCO/BFO. The higher leakage current in the bilayer arises due to hetero-valency of Fe ions, which causes the electronic instability in BFO layer of bilayer sample, consistent with the previous studies [4-6 ].

Figure S3(b) demonstrates the room temperature P-E loop behaviour of the single layer of BFO and bilayer of LCO/BFO measured at frequency value of 100 Hz. The remnant polarization values for BFO and LCO/BFO films are recorded as 48.0 and 31.5 $\mu C/cm^2$ respectively. This indicates that the bilayer also reveals the ferroelectric property, though the polarization value is reduced, which is ascribed to the hetero-valency of Fe ions induced large leakage current for the bilayer sample compared to the single layer sample. The remnant polarisation value at 100 Hz decreases to 18.8 $\mu C/cm^2$ for the sample LCO/BFO at 30 K [figure S3(c)]. This indicates that there is further increment occurring in the leakage current as we move towards low temperature region. It should be noted here that the earlier study reveals that there is the absence of any structural phase transition in BFO/STO as we move towards the lower temperature region and hence as a result, the electrical polarisation remains unaffected for single layer film of BFO as is shown in Fig. S3(d) [7-8]. Thus, it implies that the increased charge transfers from Co ion to Fe ion (as backed by the analysis of electric field dependent PES at temperature 20 K) makes more evolution of Fe 2+ charge state, which further destabilises the electronic instability due to which the polarization of LCO/BFO gets further decreases

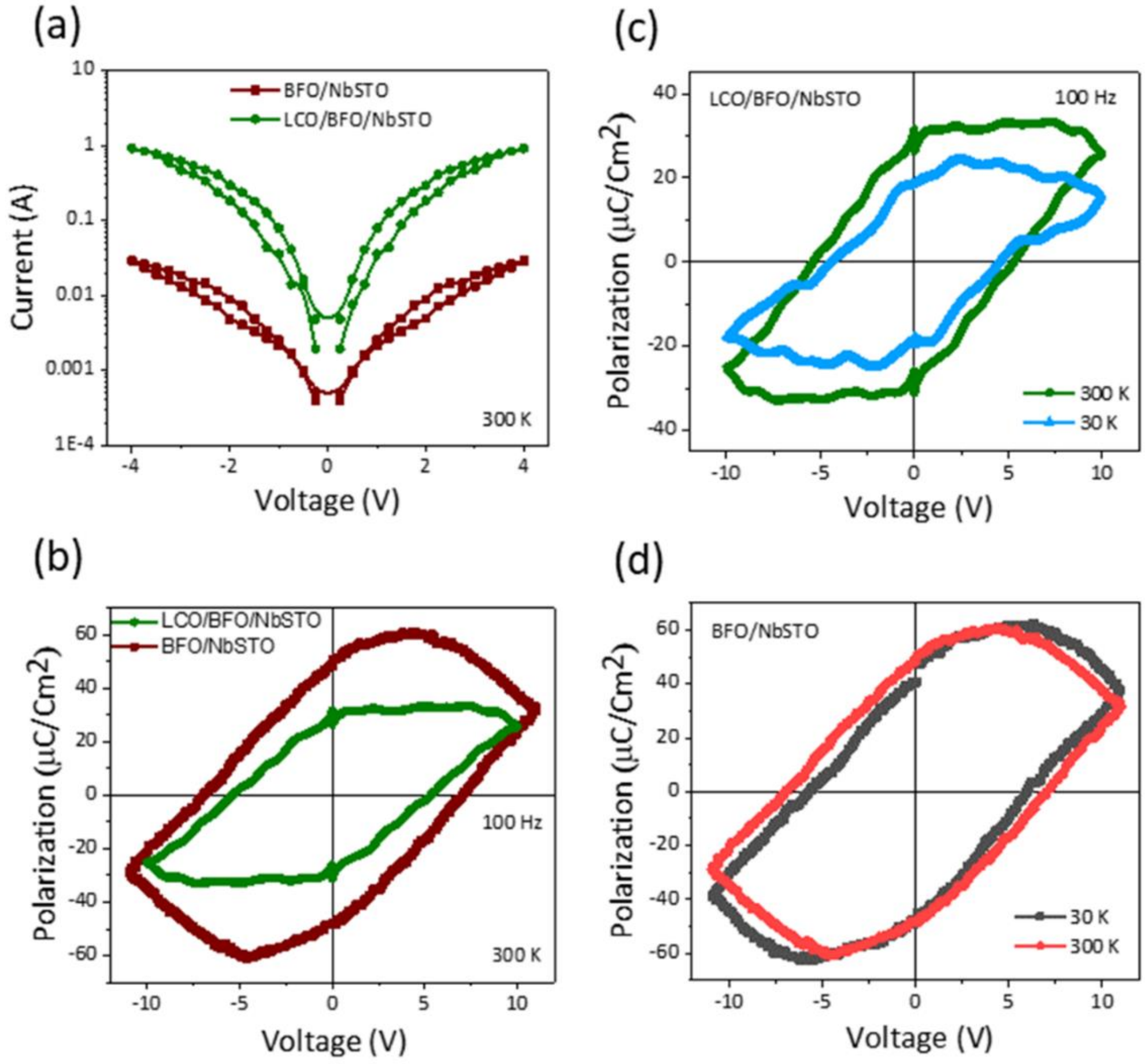

*Figure S3. (a) I-V characteristic curve taken at temperature 300 K for the single layer BFO/NbSTO and heterostructure LCO/BFO/NbSTO. (b)Polarization as a function of applied voltage/electric field for the single layer BFO/NbSTO and heterostructure LCO/BFO/NbSTO at temperature 300 K. (c)Polarization as a function of applied voltage/electric field for the heterostructure LCO/BFO/NbSTO at temperature 30 K and 300 K. (d) Polarization as a function of applied electric field for the single layer BFO/NbSTO at temperature 30 K and 300 K.*

**Magnetization measurements:**

Magnetisation vs temperature curve (figure S4(a)) of single layer of LCO on STO substrate shows ferromagnetic to paramagnetic transition at temperature around 76 K [9]. The same curve (figure S4(b)) of BFO/STO reflects a bifurcation till room temperature with a small hump around the temperature 55 K, which is attributed to the leaked oxygen molecules' in the sample chamber of the SQUID-VSM setup, which undergo paramagnetic to antiferromagnetic transition. Figure S4(c-d) shows the XMCD measurement of Co and Fe L-edge taken from single layer of LCO and BFO on STO substrate, which affirms the usual ferromagnetic and antiferromagnetic nature of the respective films in low temperature region [10-11]. As mentioned earlier, the blocking temperature for the exchange bias effect is measured to be 125 K. To acquire blocking temperature, we have carried out a field-cooled cycle measurement in both positive and negative cycles of applied magnetic field at different temperature values 5 K, 10 K, 40 K, 65 K, 100 K, 125 K (shown in figure S5(a-f)) and obtained different values of exchange bias field shift using the formula mentioned earlier [12].

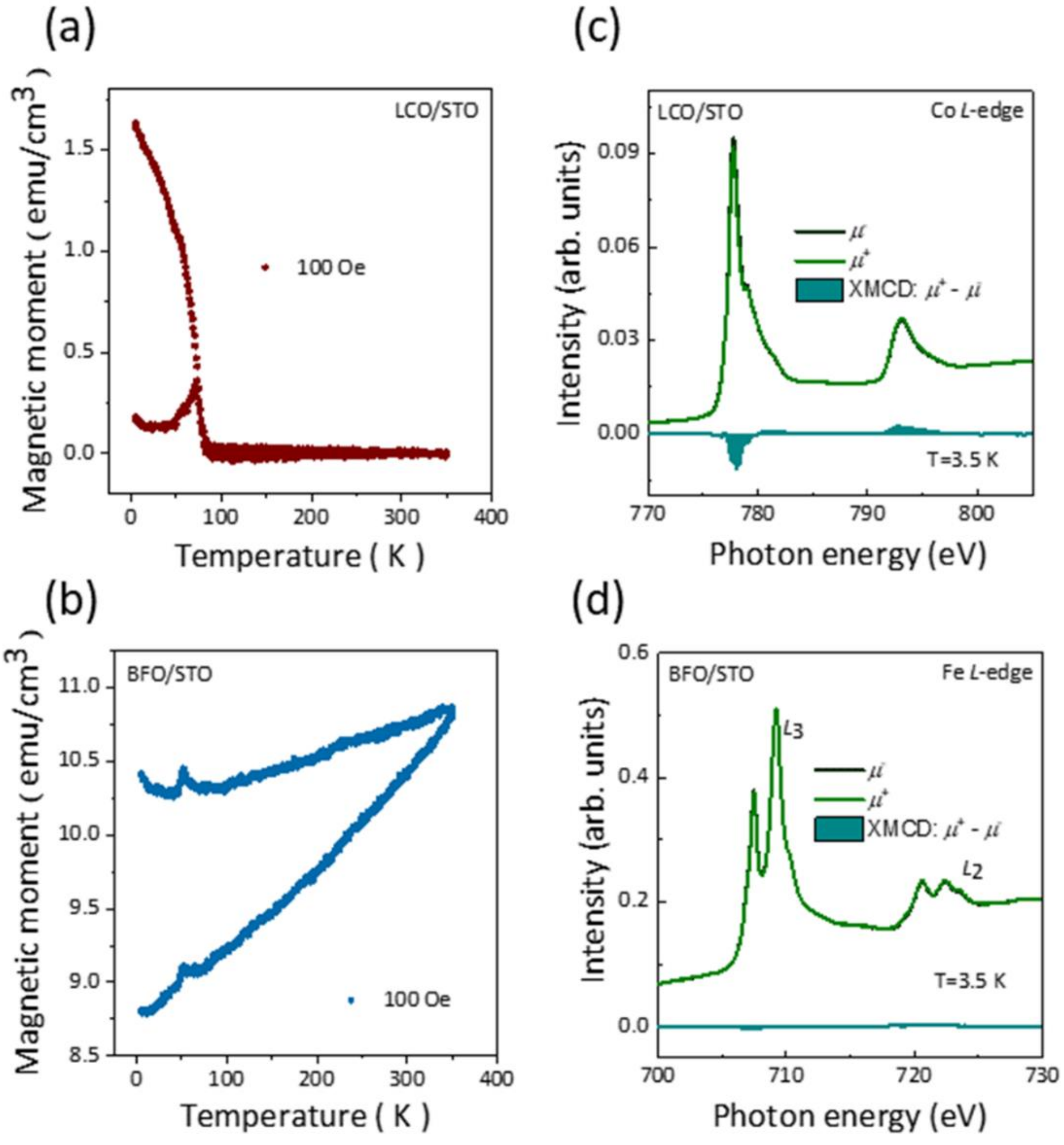

*S4. (a) and (b) M-T curve of a single layer of LCO and BFO grown on STO (001) substrate. XMCD analysis of (c) Co L-edge for the film LCO/STO and (d) Fe L-edge for the film BFO/STO at a temperature of 3.5 K.*

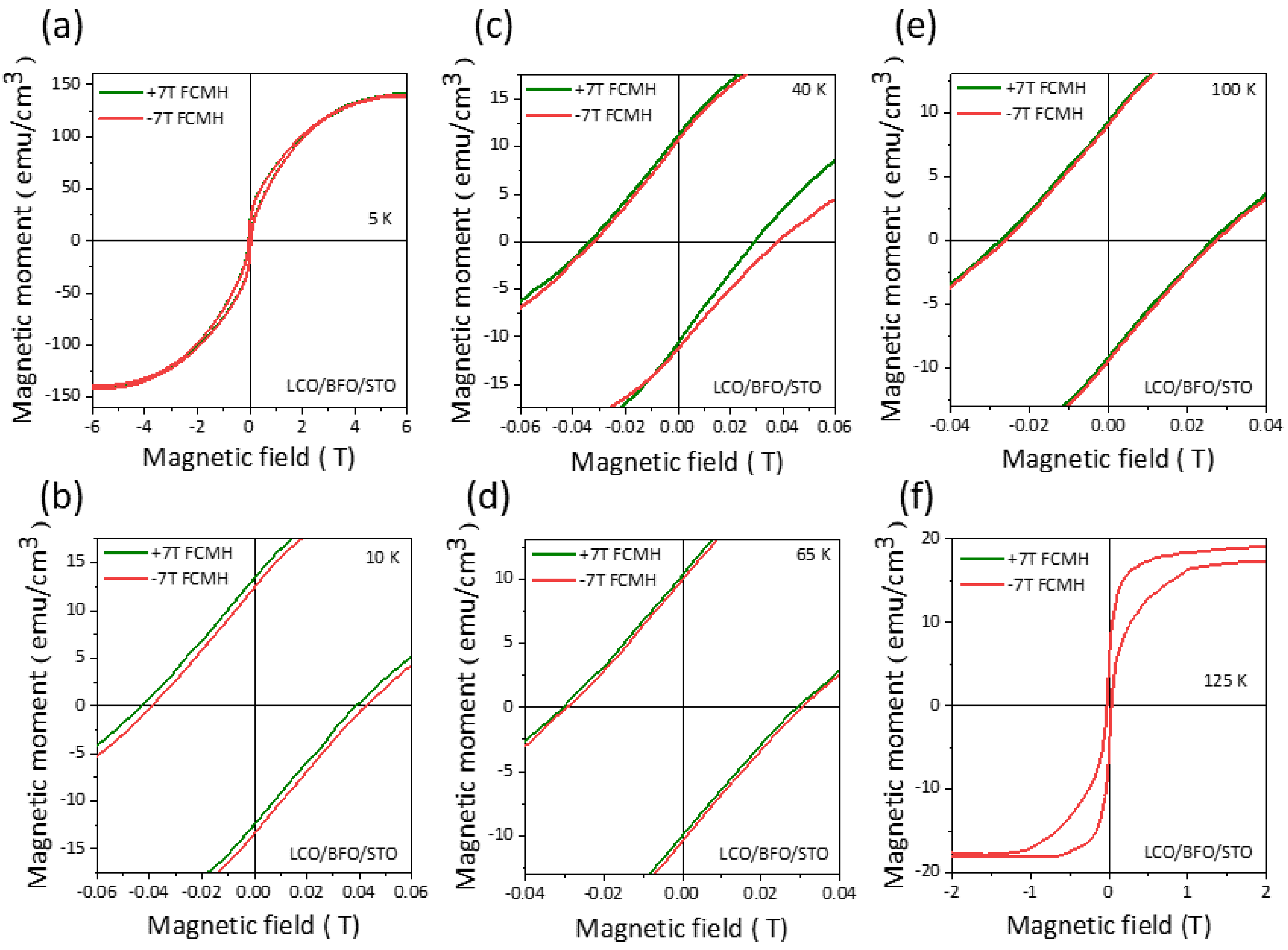

*Figure S5. (a-f) M-H plot taken at FC cycles at a temperature of 5 K,10 K, 40 K, 65 K, 100 K, 125 K for the film LCO/BFO/STO.*

## Experimental details:

### Thin film growth:

Bilayer thin film of LCO/BFO is grown on $SrTiO_3$ (001) (STO) and Nb doped STO (001) substrates by using pulsed laser deposition technique. A KrF excimer laser source is used in to ablate the dense pellet of BFO and LCO synthesized by solid state reaction. The parameter values during the formation of the bilayer sample are as follows: substrate temperature 700 $^{o}$C, oxygen partial pressure 230 mTorr, laser fluence at the target pellet 1.3 J/cm$^2$ and the distance between the target and the substrate 4.5 cm. After deposition, the samples are annealed for 10 minutes at temperature 700 $^{o}$C and then cooled down to room temperature at a rate of 10 $^{o}$C/min.

### Structural characterization:

The structural characterization was carried out by D2 Phaser x-ray diffractometer using Cu $K_{\alpha}$ source ($\lambda$=1.5046 Å). For cross-sectional imaging, the specimen was prepared using a FIB-Helios 5 UX dual-beam system. The specimen was subsequently imaged in HAADF-STEM mode using a Thermo Fisher Scientific TITAN Themis 300 transmission electron microscope equipped with a Cs aberration corrector and operated at an accelerating voltage of 300 kV. The thickness of the heterostructure was calibrated by performing x-ray reflectivity measurements (Bruker D8 diffractometer).

### Electronic characterization:

To probe the electronic state and orbital polarization, we have carried out XAS and XLD measurements respectively at soft x-ray absorption spectroscopy beamline, BL-01, Indus-2, RRCAT, India.

### Magnetic characterization:

Element-sensitive XMCD measurements were performed at BOREAS beamline, ALBA synchrotron, Barcelona, Spain, at the Co and Fe $L_{3,2}$-edges in surface-sensitive total electron yield (TEY) modes at 3.5 K at grazing incidence (20$^{o}$ with respect to the film's surface normal) under a 6 T magnetic field. The process was carried out under a background vacuum of $8 \times 10^{-9}$ Torr. The normalization of spectra was performed so that the $L_3$ pre-edge spectral region was set to zero, in addition to the post $L_2$ edge region set to one. The magnetic measurements were performed using SQUID-VSM-7T (Quantum Inc., USA).

### Electrical characterization:

The dielectric polarization of the samples was measured with PE loop tracer by following positive up negative down (PUND) pulsed sequence. For electric field dependent Photoemission Spectroscopy (PES) and PE loop measurements, a 2 nm thick gold pad was deposited on LCO layer by using an ex-situ table top coater with hard masking. Operando bias-dependent HAXPES measurements were conducted at the P22 beamline of PETRA III, DESY (Hamburg, Germany) [13] at 20 K at 6 keV of incidence energy.